\documentclass[aps,preprint,11pt,nofootinbib,groupedaddress,preprintnumbers,superscriptaddress]{revtex4-1}

\usepackage{amssymb,amsfonts,amsmath,bm,cases,empheq}
\usepackage
{graphicx}
\usepackage{color}

\usepackage
{hyperref}
\hypersetup{%
 setpagesize=false,
 bookmarksnumbered=true,
 bookmarksopen=true,
 colorlinks=true,
 linkcolor=blue,
 citecolor=red}

\begin{document}
\title{
{\Large
Observability of the innermost stable circular orbit\\ 
in a near-extremal Kerr black hole}}
\author{{\large Takahisa Igata}}
\email{igata@rikkyo.ac.jp}
\affiliation{Department of Physics, Rikkyo University, Toshima, Tokyo 171-8501, Japan}
\author{{\large Keisuke Nakashi}}
\email{nakashi@rikkyo.ac.jp}
\affiliation{Department of Physics, Rikkyo University, Toshima, Tokyo 171-8501, Japan}
\author{{\large Kota Ogasawara}}
\email{kota@tap.scphys.kyoto-u.ac.jp}
\affiliation{Theoretical Astrophysics Group, Department of Physics, Kyoto University, Kyoto 606-8502, Japan}
\date{\today}
\preprint{RUP-19-28}

\begin{abstract}
We consider the escape probability of a photon emitted 
from the innermost stable circular orbit (ISCO) 
of a rapidly rotating black hole. 
As an isotropically emitting light source on a circular orbit 
reduces its orbital radius, the escape probability of a photon emitted from it 
decreases monotonically.
The escape probability evaluated at the ISCO also decreases monotonically 
as the black hole spin increases.
When the dimensionless Kerr parameter $a$ is at the Thorne limit $a=0.998$, 
the escape probability from the ISCO is $58.8\%$. 
In the extremal case $a=1$, even if the orbital radius of the light source is 
arbitrarily close to the ISCO radius, 
which coincides with the horizon radius, 
the escape probability remains at $54.6\%$. 
We also show that such photons that have escaped from the vicinity of the horizon 
reach infinity with sufficient energy to be potentially observed
because Doppler blueshift due to relativistic beaming can overcome the gravitational redshift. 
Our findings indicate that signs of the near-horizon physics of a rapidly rotating black hole will be detectable on the edge of its shadow.
\end{abstract}
\maketitle

\section{Introduction}
\label{sec:1}

The shadow structure of the M87 galactic center (M87$^*$) discovered by 
the Event Horizon Telescope Collaboration~\cite{Akiyama:2019cqa,Akiyama:2019brx,Akiyama:2019sww,Akiyama:2019bqs,Akiyama:2019fyp,Akiyama:2019eap} 
suggests that a supermassive black hole may exist there. 
However, the possibility that the center is a horizonless compact object 
has not yet been ruled out~\cite{Mizuno:2018lxz,Akiyama:2019fyp,Cardoso:2019rvt}. 
If M87$^*$ is a black hole, 
while it is essential to observe the near-horizon region 
to distinguish it from black hole mimickers, 
because nothing can escape from the horizon, photons emitted from its vicinity 
seem unlikely to escape to infinity. 
Essentially, we can hardly perceive the scale of the gravitational radius of a black hole, 
and although we must continue to make observations to investigate physical phenomena 
at the horizon scale to identify the center, 
the simple perspective indicates the difficulty of further exploring 
the features of the central object.

However, according to recent research progress, 
this difficulty can be overcome.
One of the targets that has an advantage 
in the observability of the near-horizon region is a rapidly rotating black hole. 
It has been shown that some observable features such as bright border of 
the shadow edge appear in a near-extremal Kerr black hole~\cite{Lupsasca:2017exc,Igata:2019pgb};
more recently, it was explicitly shown that in the extremal Kerr spacetime, 
the escape probability of a photon from a source at rest
with respect to a locally nonrotating observer is $29.1\%$ 
in the limit as the emission point approaches 
the horizon~\cite{Ogasawara:2016yfk,Ogasawara:2019mir}.%
\footnote{Ratio of photons trapped by a black hole was also discussed in Ref.~\cite{Takahashi:2010ai}.}
Furthermore, the probability becomes zero in the same limit in the subextremal case, 
but in the near-extremal case, about $30\%$ is still achieved just before the horizon. 
These facts suggest that the vicinity of a rapidly rotating black hole is 
more visible than that of a slowly rotating one. 
Because M87$^*$ is generally expected to have relatively high spin~\cite{Doeleman:2012zc} 
and may be a rapidly rotating black hole~\cite{Li:2009,Feng:2017vba}, 
it is worth pursuing phenomena near the horizon of a near-extremal Kerr black hole.

In general, a light source near a black hole has proper motion relative to the horizon, 
which affects its photon escape probability and the initial energy injection to a photon. 
In particular, sources moving on stable circular orbits often exist in nature and further 
provide natural initial conditions for optically observable 
phenomena~\cite{Cunningham:1973,Cunningham:1975zz}. 
For instance, the apparent shape of the innermost edge of a standard accretion disk surrounding a black hole appears as a closed curve~\cite{Luminet:1979,Fukue:1988,Falcke:1999pj,Takahashi:2004xh,Kawashima:2019ljv,Gralla:2019drh}.
Since the innermost edge of an accretion disk 
is often identified with the innermost stable circular orbit (ISCO)~\cite{Novikov:1973kta}, 
it is important to reveal whether or not photons emitted 
from the ISCO can reach infinity with sufficient energy to be observed.
In particular, the ISCO radius coincides with the horizon radius 
in the extremal Kerr spacetime~\cite{Bardeen:1972fi,Kapec:2019hro}, 
so that the observability of the ISCO in a rapidly rotating black hole 
immediately implies that of the near-horizon region. 
The appearance of an isotropically emitting point source on a circular orbit 
in the near horizon of a rapidly rotating black hole 
has been investigated~\cite{Gralla:2015rpa,Gralla:2017ufe}, 
and it was clarified that the point source moves on the so-called NHEKline on the photon ring.

One of the additional and specific features of a source on a stable circular orbit 
in the extremal Kerr spacetime is that, 
as its orbital radius approaches the ISCO (i.e., the horizon), 
the relative velocity to the extremal horizon increases and 
can get arbitrarily close to half of the speed of light~\cite{Bardeen:1972fi}.
Emission from this yields the boost of each photon and 
photon concentration in the forward direction of the source by the relativistic beaming. 
Therefore, it can be expected that
the escape probability of a photon emitted 
from the ISCO of a rapidly rotating black hole 
becomes relatively large owing to the boost effect.

Understanding this phenomenon is important for observing certain features 
in the vicinity of the horizon. 
The purpose of this paper is to clarify how many photons emitted 
from a light source circularly orbiting 
a black hole can escape to infinity. 
In particular, 
when evaluating the escape probability of a photon 
from the ISCO in (near) extremal Kerr spacetimes, 
we demonstrate the possibility of observing the vicinity of the horizon from a distance. 
Furthermore, we also discuss the frequency shift of photons to confirm 
whether such photons reach infinity with sufficient energy.

This paper is organized as follows.
In Sec.~\ref{sec:2}, we derive the conditions for photon escape 
from the vicinity of the Kerr black hole horizon to infinity. 
In Sec.~\ref{sec:3}, by using the results in the previous section, 
we define and show the escape cone of a photon emitted from a light source 
orbiting a black hole. Furthermore, we evaluate the escape probability 
under the assumption that the emission is isotropic. 
We also discuss the redshift for a photon observed at infinity. 
In Sec.~\ref{sec:4}, we present a summary and discussions. 
Throughout this paper, we use units in which $G=1$ and $c=1$.

\section{Escape of a photon from the vicinity of the Kerr black hole horizon}
\label{sec:2}
The Kerr metric in the Boyer--Lindquist coordinates is given by
\begin{align}
&g_{\mu\nu}\:\!\mathrm{d}x^\mu \mathrm{d}x^\nu
=-\frac{\Sigma\Delta}{A}\mathrm{d}t^2+\frac{\Sigma}{\Delta}\mathrm{d}r^2
+\Sigma \:\!\mathrm{d}\theta^2+\frac{A}{\Sigma}\sin^2\theta\left[\:\!
\mathrm{d}\varphi-\frac{a(r^2+a^2-\Delta)}{A} \mathrm{d}t\:\!\right]^2,
\label{eq:metric}
\\
&\Sigma=r^2+a^2\cos^2\theta,
\quad
\Delta=r^2+a^2-2Mr,
\quad
A=(r^2+a^2)^2-\Delta a^2 \sin^2\theta,
\end{align}
where $M$ and $a$ are mass and spin parameters, respectively. 
We assume $0<a \leq M$ 
in order to focus on a rotating black hole spacetime. 
The event horizon is located at the radius $r=r_{\mathrm{h}}=M+\sqrt{M^2-a^2}$. 
If $a=M$, the metric describes the extremal Kerr black hole spacetime, where $r_{\mathrm{h}}=M$. 
In the following part, units in which $M=1$ are used.

\medskip

We consider photon motion in the Kerr black hole spacetime. Let $k^a$ be a null geodesic tangent. 
Because the metric admits the Killing vectors $\partial/\partial t$ and $\partial/\partial \varphi$, 
a free photon has constants of motion 
$k_t=k_a (\partial/\partial t)^a$ and $k_\varphi=k_a (\partial/\partial \varphi)^a$. 
Furthermore, the metric also admits the Killing tensor
\begin{align}
K_{ab}=\Sigma^2 (\mathrm{d}\theta)_a(\mathrm{d} \theta)_b
+\sin^2\theta \left[\:\!
(r^2+a^2) (\mathrm{d}\varphi)_a-a \:\!(\mathrm{d}t)_a
\:\!\right]\left[\:\!
(r^2+a^2) (\mathrm{d}\varphi)_b-a \:\!(\mathrm{d}t)_b
\:\!\right]-a^2\cos^2\theta g_{ab}, 
\end{align}
such that a photon has a quadratic constant of motion in $k^a$ given by
\begin{align}
Q=K_{ab}k^a k^b-(k_\varphi+a k_t)^2. 
\end{align}
We assume positivity of 
photon energy (i.e., $-k_t>0$) because we only focus on photons reaching infinity. 
Here, we introduce the impact parameters,
\begin{align}
b=-\frac{k_\varphi}{k_t}, \quad 
q=\frac{Q}{k_t^2}.
\end{align}
Rescaling the tangent $k^a$ by $-k_t$ (i.e., $-k^a/k_t \to k^a$), we rewrite 
the components of $k^a$ in terms of $b$ and $q$ as follows:
\begin{align}
&k^t=\dot{t}=\frac{1}{\Sigma} \left[\:\!
a\left(b-a\sin^2\theta\right)
+\frac{r^2+a^2}{\Delta}\left(
r^2+a^2-ab 
\right)
\:\!\right],
\\
\label{eq:kr}
&k^r=\dot{r}=\frac{\sigma_r}{\Sigma} \sqrt{R},
\\
\label{eq:ktheta}
&k^\theta=\dot{\theta}=\frac{\sigma_\theta}{\Sigma} \sqrt{\Theta},
\\
&k^\varphi=\dot{\varphi}=\frac{1}{\Sigma}\left[\:\!
\frac{b}{\sin^2\theta}-a+\frac{a}{\Delta} \left(r^2+a^2-a\:\!b\right)
\:\!\right],
\end{align}
where $\sigma_r$, $\sigma_\theta=\pm$, the dot denotes 
the derivative with respect to an affine parameter, and
\begin{align}
R&=\left(r^2+a^2-a\:\!b\right)^2-\Delta \left[\:\!
q+(b-a)^2\:\!\right],
\\
\label{eq:Theta}
\Theta&=q-b^2\cot^2\theta+a^2\cos^2\theta.
\end{align}

\begin{figure}[t!]
\centering
 \includegraphics[width=14cm,clip]{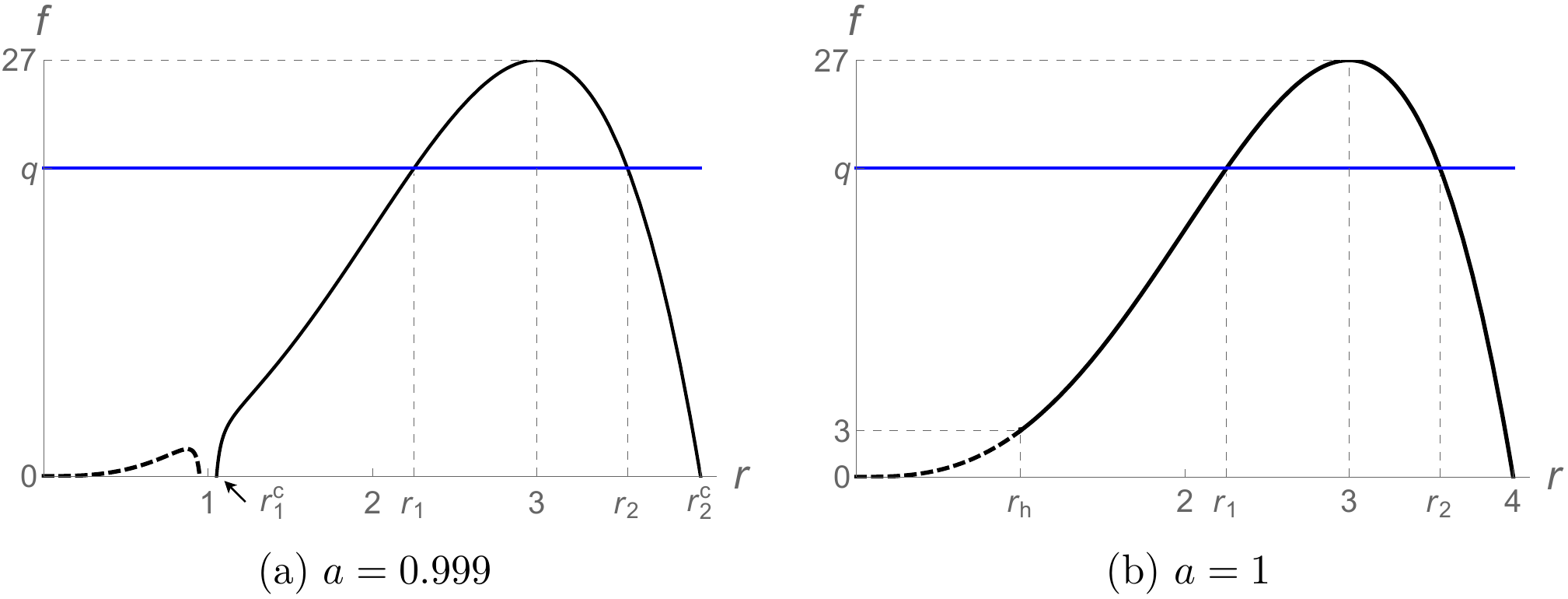}
 \caption{
Relation between the impact parameter $q$ and the radii $r_i$ ($i=1, 2$) of extremum points of the effective potentials $b_i(r)$. The function $f$ is shown by black lines, 
which are solid outside the horizon and dashed inside 
it. The left panel~(a) is a subextremal case ($a=0.999$), 
and the right panel~(b) is the extremal case ($a=1$).}
 \label{fig:f}
\end{figure}

We review conditions for a photon escaping 
from the vicinity of the horizon to infinity~\cite{Ogasawara:2019mir}.
To derive them, we analyze the equation of photon radial motion~\eqref{eq:kr}, which is rewritten as 
\begin{align}
\dot{r}^2+\frac{r\:\!(r-2)}{\Sigma^2}
(b-b_1)(b-b_2)
=0, 
\end{align}
where 
\begin{align}
b_1(r)&=\frac{-2\:\!a\:\!r +\left[\:\!r\Delta (r^3-q\:\!r+2\:\!q)\:\!\right]^{1/2}}{r(r-2)},
\\
b_2(r)&=\frac{-2\:\!a\:\!r -\left[\:\!r\Delta (r^3-q\:\!r+2\:\!q)\:\!\right]^{1/2}}{r(r-2)}.
\end{align}
We call $b_i$ ($i=1, 2$) the effective potentials for photon radial motion. 
The allowed parameter range of $b$ for a positive energy photon is 
\begin{align}
&b\leq b_1\ \ 
\mathrm{for} \ \  r_{\mathrm{h}} < r<2,
\\
&b_2\leq b\leq b_1 \ \ \mathrm{for} \ \ r\geq 2.
\end{align}
Since the escape of a photon is governed by the potential barrier, 
the extremum points of $b_i$ ($i=1, 2$) 
should be known. 
Solving $b_i'(r)=0$ for $q$, we obtain a common equation
\begin{align}
\label{eq:q=f}
q&=f(r)\equiv\frac{r^2}{a^2} \left[\:\!
-\frac{4(1-a^2) r}{(r-1)^2}
+3+(3-r)(r-1)
\:\!\right]. 
\end{align}
It is sufficient to investigate the region where $f(r)\geq 0$ 
because 
the light source we discuss is located on the equatorial plane,
and therefore, from 
Eqs.~\eqref{eq:ktheta} and \eqref{eq:Theta}, a photon must initially satisfy $\Theta=q\geq0$.
Figure~\ref{fig:f}(a) shows a typical shape of $f(r)$ 
in the case of subextremal Kerr spacetimes (i.e., $0<a<1$). 
This figure shows that Eq.~\eqref{eq:q=f} has two roots $r_1$, $r_2$ outside the horizon, 
which are restricted in the range 
\begin{align}
\label{eq:rorder}
&r_{\mathrm{h}} < r_{1}^\mathrm{c}\leq r_1 \leq 3 \leq r_2 \leq r_{2}^\mathrm{c},
\end{align}
where $r=3$ is a local maximum point of $f$, and
\begin{align}
&r_{1}^\mathrm{c}=2+2\cos \left[\:\!
\frac{2}{3} \arccos(a)-\frac{2\pi}{3}
\:\!\right],
\\
&r_{2}^\mathrm{c}=2+2\cos \left[\:\!
\frac{2}{3} \arccos(a)
\:\!\right]
\end{align}
are radii of circular photon orbits solving the equation $f=0$ (i.e., $q=0$). 
When a photon stays at the top of an extremum point of $b_1$ ($b_2$), 
the orbit with constant radius $r=r_1$ ($r=r_2$) 
is called the spherical photon orbit. Then the photon must have 
\begin{align}
b=b_i^{\mathrm{s}}\equiv
\frac{2\:\!(1-a^2)}{a(r_i-1)}
-\frac{(r_i-1)^2}{a}+\frac{3}{a}-a.
\end{align}

Next, we focus on the case of the extremal Kerr spacetime (i.e., $a=1$), 
for which the first term in $f$ vanishes 
such that Eq.~\eqref{eq:q=f} reduces to
\begin{align}
q=r^3(4-r). 
\end{align}
Figure~\ref{fig:f}(b) shows the shape of $f$ in the extremal case (i.e., $a=1$). 
Contrary to the subextremal case,
the number of roots of Eq.~\eqref{eq:q=f} depends on $q$. 
Outside the horizon, 
there exists a single root $r_2$ for $0\leq q\leq 3$, while 
there exist two roots $r_1$, $r_2$ outside the horizon for 
$3<q<27$, where the boundary values of the ranges come from 
$f(r_{\mathrm{h}})=3$ and $f(3)=27$.

\medskip

Let us consider escape conditions for a photon in terms of $(b, q)$. 
Classifying the radial position $r$ of a light source 
based on relations with the extremum point $r_1$ and the horizon $r_{\mathrm{h}}$, we clarify the parameter range of $(b, q)$ for a photon that can escape from the vicinity of the horizon to infinity.
In the following part, it is assumed that $r<3$.

\begin{figure}[t!]
\centering
 \includegraphics[width=16cm,clip]{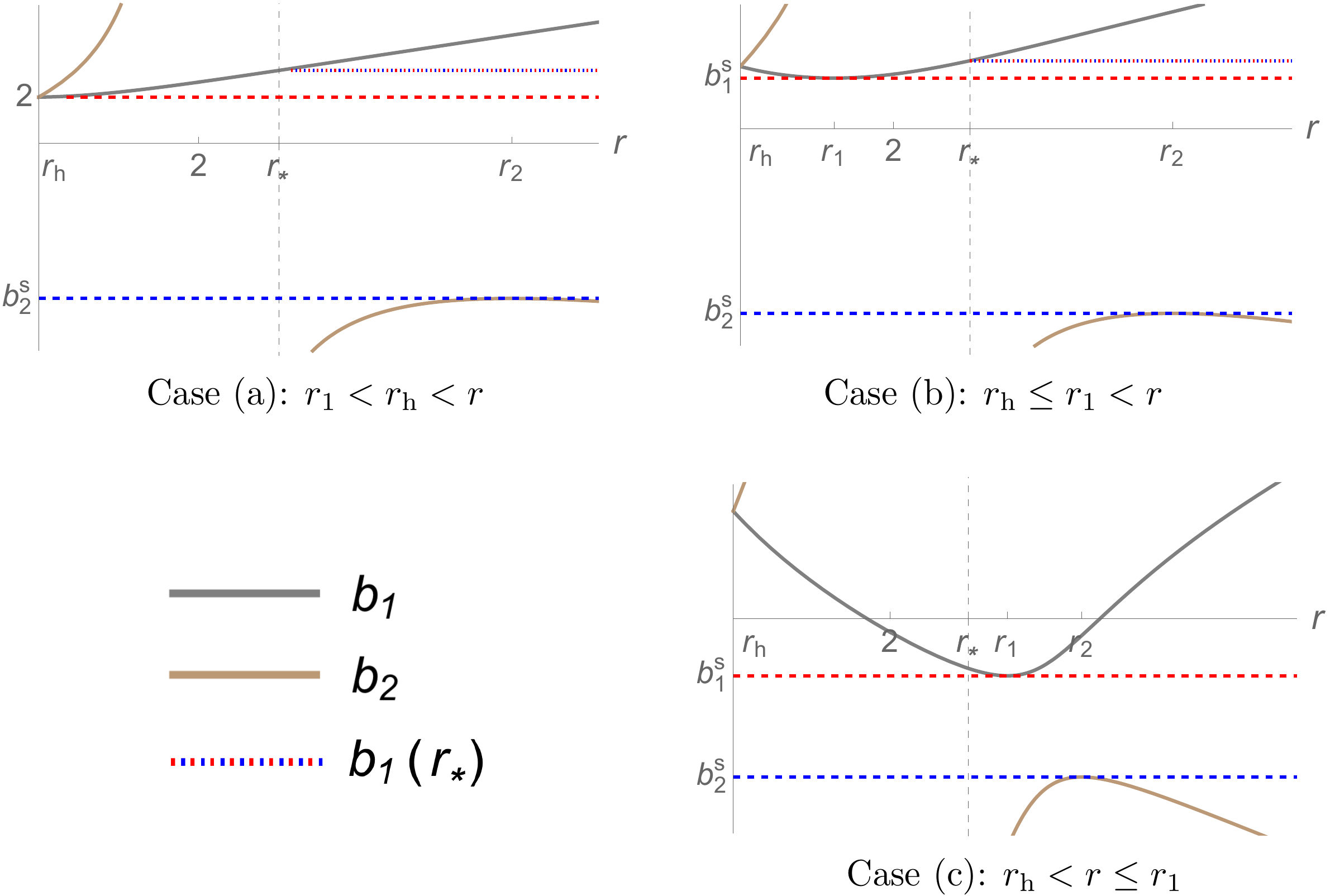}
 \caption{Typical shapes of the effective potentials $b_i$ $(i=1, 2)$ and parameter ranges of $b$ for escaping photons. The radius $r_*$ indicates the radial position of the source. 
}
 \label{fig:case}
\end{figure}

\nopagebreak[1]

\textit{Case~a.---}We consider the case $r_1<r_{\mathrm{h}}<r$, 
which we define as Case~a. 
Note that this inequality appears only for the extremal case because 
$r_1>r_\mathrm{h}$ always holds for the subextremal case [see Fig.~\ref{fig:f} 
and Eq.~\eqref{eq:rorder}]. The first inequality $r_1<r_{\mathrm{h}}$ leads to 
$0\leq q <3$.
In this range of $q$, 
we have typical plots of $b_1$, $b_2$ and marginal parameter values of 
$b$, as
shown in Fig.~\ref{fig:case}(a). 
From this plot, the allowed range of $b$ for escape
can be read as follows:
if $\sigma_r=+$ (i.e., radially outward emission), 
a photon with 
$b_2^{\mathrm{s}}<b\leq b_1$ 
(i.e., the bounded range between the blue lines) can escape to infinity. 
Even if $\sigma_r=-$ (i.e., radially inward emission), 
a photon with 
$2<b< b_1$ 
(i.e., the bounded range between the red lines) can do so. 
Hence, we find that the marginal parameter values of $b$ are given by
$(\sigma_r, b)=(+, b_2^{\mathrm{s}}), \ (-, 2)$.

\textit{Case~b.---}We consider the case 
$r_{\mathrm{h}}\leq r_1 <r$,
which we define as Case~b. 
The corresponding range of $q$ is 
$3\leq q < f$ 
for $a=1$ 
and 
$0\leq q < f$
for 
$0<a<1$.
Figure~\ref{fig:case}(b) shows typical plots of $b_i$ in these ranges of $q$. 
If $\sigma_r=+$, a photon with 
$b_2^{\mathrm{s}} < b \leq b_1$ 
(i.e., between the blue lines) can escape infinity, while if $\sigma_r=-$, a photon with 
$b_1^{\mathrm{s}}<b<b_1$ (i.e., between the red lines) can do so. 
Hence, we find that marginal parameter values of $b$ are given by
$(\sigma_r, b)=(+, b_2^{\mathrm{s}}), (-, b_1^{\mathrm{s}})$.

\textit{Case~c.}---We consider the case 
$r_{\mathrm{h}}<r\leq r_1$,
which we define as Case~c. 
Then the parameter range of $q$ is restricted to $f\leq q<27$.
Figure~\ref{fig:case}(c) shows typical plots of $b_i$ in this ranges. 
Only if $\sigma_r=+$, a photon can escape to infinity, and it must then have
$b_2^{\mathrm{s}}<b<b_1^\mathrm{s}$ (i.e., between the red and blue lines).
Hence, we find that marginal parameter values $b$ are given by 
$(\sigma_r, b)=(+, b_1^{\mathrm{s}}), (+, b_2^{\mathrm{s}})$.

The allowed parameter values of $(b, q)$ for escape from the vicinity of the horizon are summarized in Tables~\ref{table:extremal} and \ref{table:nonextremal}. 
\begin{table}[t!]
\begin{tabular}{lllll}
\hline \hline
\multicolumn{1}{c}{Cases~~~~~~~}&\multicolumn{1}{c}{$q$~~~~~~~~~~~~~}&\multicolumn{1}{c}{$b$ ($\sigma_r=+$)~~~~~~~~~~} &\multicolumn{1}{c}{$b$ ($\sigma_r=-$)~~~~~~~~}&\multicolumn{1}{c}{Marginal pairs of $(\sigma_r, b)$~~}
 \\
 \hline
\multicolumn{1}{l}{(a) $r_1<r_{\mathrm{h}}<r$}
~~~~~~~&
$0\leq q<3$
~~~~~~~~~~~
&
$b_2^{\mathrm{s}}<b\leq b_1$
~~~~~~~
&
$2<b< b_1$
~~~~
&$(+, b_2^{\mathrm{s}})$ and $(-, 2)$
\\
\multicolumn{1}{l}{(b) $r_{\mathrm{h}} \leq r_1 < r$}&$3\leq q< f
$&$b_2^{\mathrm{s}}<b\leq b_1
$&$b_1^{\mathrm{s}}<b< b_1
$&$(+, b_2^{\mathrm{s}})$ and $(-, b_1^{\mathrm{s}})$
\\
\multicolumn{1}{l}{(c) $r_{\mathrm{h}}<r
\leq r_1$}&$f
\leq q<27$&$b_2^{\mathrm{s}}<b<b_1^{\mathrm{s}}$&\multicolumn{1}{c}{n/a}~~~~~~~~&$(+, b_2^{\mathrm{s}})$ and $(+, b_1^{\mathrm{s}})$
\\
\hline 
\hline
\end{tabular}
\caption{($a=1$)
Allowed parameter values of $(b, q)$ for escape from the vicinity of the horizon in the extremal Kerr spacetime.}
\label{table:extremal}
\end{table}
\begin{table}[t!]
\begin{tabular}{cllll}
\hline \hline
\multicolumn{1}{c}{Cases~~~~~~~}&\multicolumn{1}{c}{$q$~~~~~~~~~}&\multicolumn{1}{c}{$b$ ($\sigma_r=+$)~~~~~~~~~} &\multicolumn{1}{c}{$b$ ($\sigma_r=-$)~~~~~~~}&\multicolumn{1}{c}{Marginal pairs of $(\sigma_r, b)$}
 \\
 \hline
\multicolumn{1}{l}{(b) $r_{\mathrm{h}} \leq r_1 < r
$}
~~~~~~~&
$0\leq q< f
$
~~~~~~~
&
$b_2^{\mathrm{s}}<b\leq b_1
$
~~~~~~~&
$b_1^{\mathrm{s}}<b< b_1
$
~~~~
&$(+, b_2^{\mathrm{s}})$ and $(-, b_1^{\mathrm{s}})$
\\
\multicolumn{1}{l}{(c) $r_{\mathrm{h}}<r
\leq r_1$}&$f
\leq q<27$&$b_2^{\mathrm{s}}<b<b_1^{\mathrm{s}}$&\multicolumn{1}{c}{n/a}~~~~~~~~&$(+, b_2^{\mathrm{s}})$ and $(+, b_1^{\mathrm{s}})$
\\
\hline 
\hline
\end{tabular}
\caption{($0<a<1$)
Allowed parameter values of $(b, q)$ for escape from the vicinity of the horizon in a subextremal Kerr black hole spacetime.
}
\label{table:nonextremal}
\end{table}

\section{Escape cone, escape probability, and redshift}
\label{sec:3}

We focus on a light source circularly orbiting a rotating black hole 
on the equatorial plane $\theta=\pi/2$. 
The energy $E$ and angular momentum $L$ of a source in a prograde circular orbit 
of radius $r$
are given by~\cite{Bardeen:1972fi}
\begin{align}
E
&=\frac{r^{3/2}-2\:\!r^{1/2}+a}{r^{3/4} (r^{3/2}-3 \:\!r^{1/2}+2 \:\!a )^{1/2}},
\\
L&=\frac{r^2-2\:\!a\:\!r^{1/2}+a^2}{r^{3/4}(
r^{3/2}-3\:\!r^{1/2}+2\:\!a
)^{1/2}}. 
\end{align}
We restrict the orbital radius $r$ of the source in the range $r\geq r_{\mathrm{I}}$, 
where $r_{\mathrm{I}}$ is the ISCO radius,
\begin{align}
r_{\mathrm{I}}&=3+Z_2-\left[\:\!
(3-Z_1)(3+Z_1+2 Z_2)
\:\!\right]^{1/2},
\\
Z_1&=1+(1-a^2)^{1/3}\left[\:\!
(1+a)^{1/3}+(1-a)^{1/3}
\:\!\right],
\\
Z_2&=(3\:\!a^2+Z_1^2)^{1/2}.
\end{align}
Note that $r_{\mathrm{I}}$ approaches $r_{\mathrm{h}}$ as $a$ approaches the extremal value $1$. 
In the extremal case, $r_{\mathrm{I}}$ coincides with the horizon radius, i.e., $r_{\mathrm{I}}=r_{\mathrm{h}}=1$. 

To describe a photon emission from the source, 
we introduce a frame associated with its rest frame,
\begin{align}
e^{(0)}&=-E\:\!\mathrm{d}t+L\:\!\mathrm{d}\varphi,
\\
e^{(1)}&=\frac{r}{\sqrt{\Delta}}\:\!\mathrm{d}r,
\\
e^{(2)}&=r\:\!\mathrm{d}\theta,
\\
e^{(3)}&=\frac{\sqrt{\Delta}}{r^{3/4} (r^{3/2}-3\:\!r^{1/2}+2\:\!a)^{1/2}}\left[\:\!
(r^{3/2}+a)\:\!\mathrm{d}\varphi-\mathrm{d}t
\:\!\right],
\end{align}
which is a tetrad 
only on massive particle circular orbits in the equatorial plane. 
Then, 
the tetrad components of $k^a$ at the source position are 
$k^{(\mu)}=k^a e^{(\mu)}{}_a|_{\theta=\pi/2}$, where 
\begin{align}
\label{eq:k0}
k^{(0)}&=\frac{b-a-r^{3/2}}{r^{3/4} (r^{3/2}-3\:\!r^{1/2}+2\:\!a)^{1/2}},
\\
k^{(1)}&=\frac{\sigma_r}{r} \sqrt{\frac{R}{\Delta}}, 
\\
\label{eq:k2}
k^{(2)}&=\sigma_\theta \frac{\sqrt{q}}{r},
\\
\label{eq:k3}
k^{(3)}&=\frac{b E-L}{\sqrt{\Delta}}.
\end{align}

\begin{figure}[t!]
\centering
 \includegraphics[width=6cm,clip]{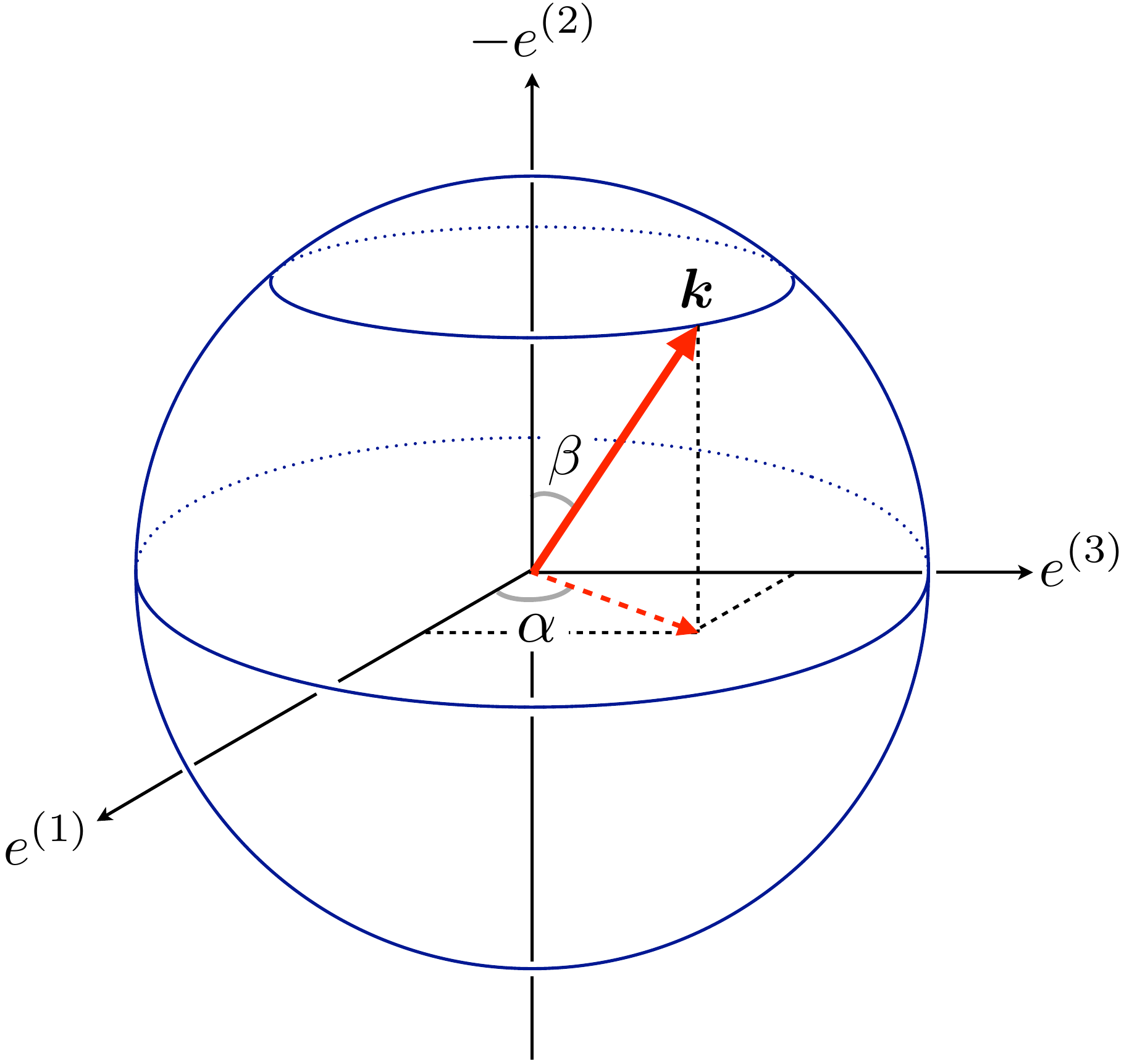}
 \caption{Definition of photon emission angles at a light source. }
 \label{fig:celestial}
\end{figure}

We parametrize the spatial direction of photon emission 
by two angle parameters $(\alpha, \beta)$ as follows:
\begin{align}
\cos \alpha \sin \beta=\frac{k^{(1)}}{k^{(0)}},
\quad
\cos \beta=-\frac{k^{(2)}}{k^{(0)}},
\quad
\sin \alpha \sin \beta=\frac{k^{(3)}}{k^{(0)}},
\end{align}
where $\beta$ is the polar angle measured from the direction $-e^{(2)}$ 
to the direction of $\bm{k}$ (projection of $k^a$ normal to $e^{(0)}$),
and $\alpha$ is the azimuthal angle measured from the direction $e^{(1)}$ 
to the projection of $\bm{k}$ on the plane spanned by $\{e^{(1)}, e^{(3)}\}$.%
\footnote{The angles $\alpha$ and $\beta$ are related to the directional cosines $\Psi$ and $\Theta$ of a beam of radiation  with respect to $\phi$-direction and $\theta$-direction in Ref.~\cite{Cunningham:1973} as 
\begin{align}
\Theta=\beta, \quad\cos\Psi=\sin\alpha \sin\beta.
\end{align}}
Figure~\ref{fig:celestial} shows the relation between $\bm{k}$ and $(\alpha, \beta)$. 
Solving these relations for $(\alpha, \beta)$ and using the null condition, we have
\begin{alignat}{3}
\cos \alpha&=\frac{k^{(1)}}{\sqrt{(k^{(1)})^2+(k^{(3)})^2}},
\quad&&
\sin \alpha=\frac{k^{(3)}}{\sqrt{(k^{(1)})^2+(k^{(3)})^2}},
\\
\cos \beta&=-\frac{k^{(2)}}{k^{(0)}},
\quad&&
\sin \beta=\frac{\sqrt{(k^{(1)})^2+(k^{(3)})^2}}{
k^{(0)}}.
\end{alignat}
As a result, a pair $(b, q)$ has a one-to-one relation
to a pair of emission angles $(\alpha, \beta)$. 
Hence, the allowed parameter ranges of $(b, q)$ 
for photon escape restrict the range of $(\alpha, \beta)$. 

\medskip

We now relate marginal parameter values for photon escape to $(\alpha, \beta)$. 
Let $S$ be the complete set of emission angles $(\alpha, \beta)$ at which 
a photon can escape to infinity. We call $S$ the escape cone of a photon.
If a photon has emission angles of the boundary values of $S$, 
it cannot escape to infinity anymore. 
We call the set of all critical emission angles $\partial S$; 
this set can be explicitly specified
in terms of marginal parameter values given 
in Tables~\ref{table:extremal} and \ref{table:nonextremal} as follows:
\begin{align}
\partial S=\bigcup_{i=1, 2} \left\{
(\alpha_i,\beta_i)\, \big|\,  0\leq q\leq 27
\right\},
\end{align}
where, in the extremal case, we have defined
\begin{subequations}
\label{def:critical_angles1}
\begin{empheq}[left={\big(\alpha_1,\beta_1\big)\equiv\empheqlbrace}]{alignat=2}
&\big(\alpha_\textrm{1(a)},\beta_{1(\mathrm{a})}\big) 
\equiv \big(\alpha,\beta \big)
\big|_{\substack{
\begin{subarray}{l}
\sigma_r=- \\ b=2
\end{subarray}}} 
&~\textrm{for} 
& ~ 0\leq q< 3,
\\[3mm]
&\big(\alpha_\textrm{1(b)},\beta_{1(\mathrm{b})}\big) 
\equiv \big(\alpha,\beta \big)
\big|_{\substack{
\begin{subarray}{l}
\sigma_r=- \\ b=b_1^{\textrm{s}}
\end{subarray}
}}
&~\textrm{for} 
& ~ 3\leq q < f,
\\[3mm]
&\big(\alpha_\textrm{1(c)},\beta_{1(\mathrm{c})}\big) 
\equiv \big(\alpha,\beta \big)
\big|_{\substack{
\begin{subarray}{l}
\sigma_r=+ \\ b=b_1^{\textrm{s}}
\end{subarray}
}} 
&~\textrm{for} 
&~ f \leq q\leq 27,
\end{empheq}
\end{subequations}
\begin{align}
\big(\alpha_2,\beta_2\big)\equiv\big(\alpha,\beta\big)\big|_{\substack{
\begin{subarray}{l}
\sigma_r=+ \\ b=b_2^{\textrm{s}} 
\end{subarray}
}} ~~\textrm{for}~~0\leq q\leq 27, 
\end{align}
and in a subextremal case, 
\begin{subequations}
\begin{empheq}[left={\big(\alpha_1,\beta_1\big)\equiv\empheqlbrace}]{alignat=2}
&\big(\alpha_{\textrm{1(b)}},\beta_{1(\mathrm{b})}\big) 
\equiv \big(\alpha,\beta \big)
\big|_{\substack{
\begin{subarray}{l}
\sigma_r=- \\ b=b_1^{\textrm{s}} 
\end{subarray}
}} 
&~ \textrm{for} 
& ~0\leq q < f,
\\[3mm]
&\big(\alpha_{\mathrm{1(c)}},\beta_{1(\mathrm{c})}\big) 
\equiv \big(\alpha,\beta \big)
\big|_{\substack{
\begin{subarray}{l}
\sigma_r=+ \\ b=b_1^{\textrm{s}} 
\end{subarray}
}} 
&~ \textrm{for} 
& ~ f \leq q\leq 27,
\end{empheq}
\end{subequations}
\begin{align}
\big(\alpha_2,\beta_2\big)\equiv \big(\alpha,\beta\big)\big|_{\substack{
\begin{subarray}{l}
\sigma_r=+ \\ b=b_2^{\textrm{s}} 
\end{subarray}
}} ~~\textrm{for}~~0\leq q\leq 27.
\end{align}

\begin{figure}[t!]
\centering
 \includegraphics[width=16.4cm,clip]{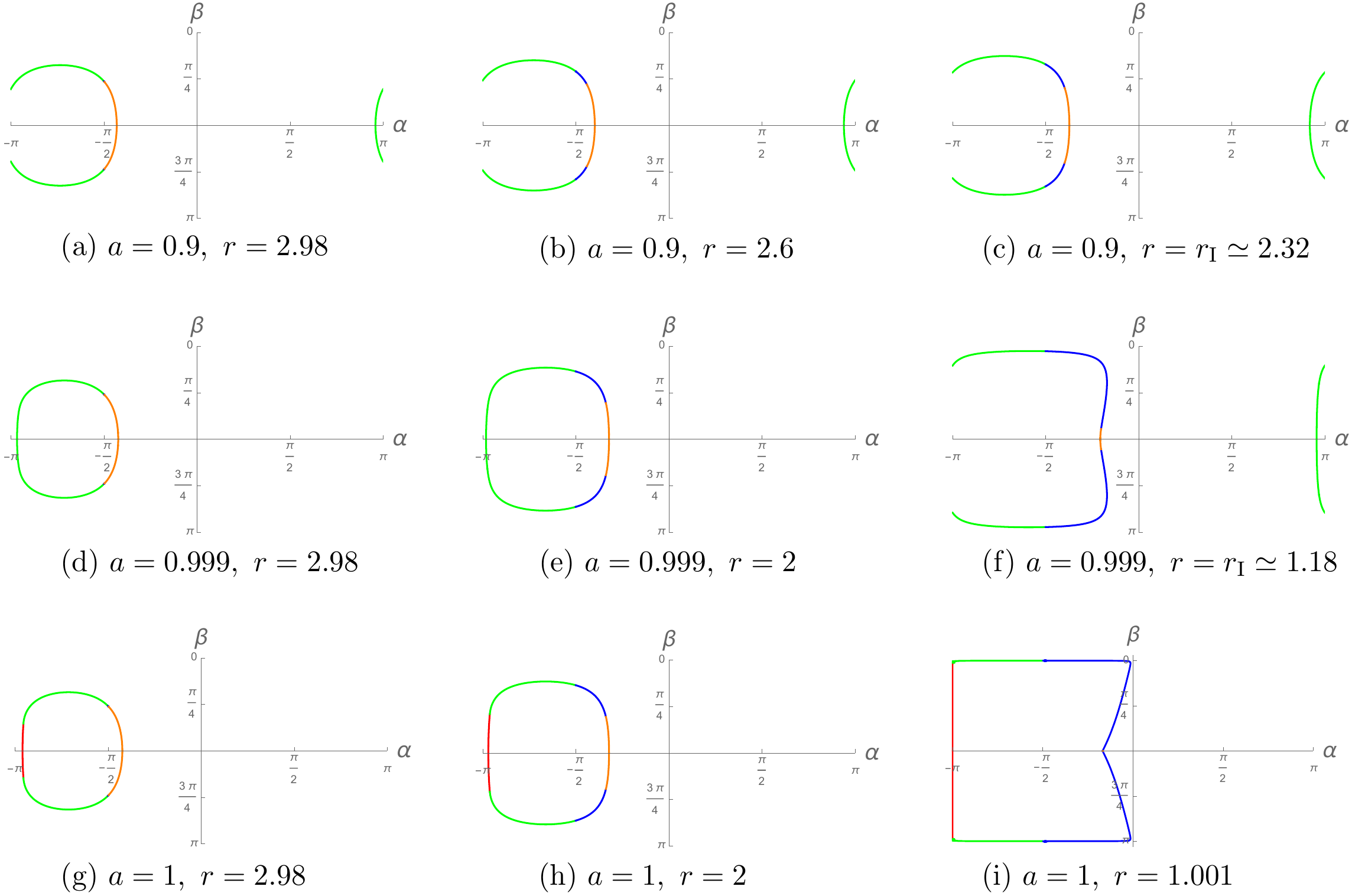}
 \caption{Escape cone of a photon emitted at a source circularly orbiting
 a rotating black hole. The first, second, and third lines show the cases $a=0.9$, $a=0.999$, and $a=1$, respectively.
The red, green, blue, and orange lines show
$(\alpha_{1(\mathrm{a})}, \beta_{1(\mathrm{a})})$, 
$(\alpha_{1(\mathrm{b})}, \beta_{1(\mathrm{b})})$,
$(\alpha_{1(\mathrm{c})}, \beta_{1(\mathrm{c})})$,
and 
$(\alpha_2, \beta_2)$, respectively.
The area containing the origin bounded by the colored lines corresponds to the escape cone of a photon.}
 \label{fig:cones}
\end{figure}

Figure~\ref{fig:cones} shows the critical emission angles for photon escape. 
The red, green, blue, and orange lines show the critical angles 
$(\alpha_{1(\mathrm{a})}, \beta_{1(\mathrm{a})})$, 
$(\alpha_{1(\mathrm{b})}, \beta_{1(\mathrm{b})})$, 
$(\alpha_{1(\mathrm{c})}, \beta_{1(\mathrm{c})})$, and 
$(\alpha_{2}, \beta_{2})$, respectively. 
Note that the region containing the coordinate origin bounded by critical angles 
corresponds to an escape cone. 
Figures~\ref{fig:cones}(a)--\ref{fig:cones}(c) show escape cones 
in the cases $r=2.98$, $2.6$, $r_{\mathrm{I}}(= 2.32\cdots)$, respectively, for $a=0.9$. 
The horizon radius is $r_{\mathrm{h}}\simeq 1.43\cdots$.
The area of the escape cones becomes smaller as 
the emission point approaches the ISCO. 
Figures~\ref{fig:cones}(d)--\ref{fig:cones}(f) show escape cones in the cases
$r=2.98$, $2$, $r_{\mathrm{I}}(= 1.18\cdots)$, respectively, for $a=0.999$. 
The horizon radius is $r_{\mathrm{h}}=1.04\cdots$. 
As is the case for $a=0.9$, 
the area of the escape cones becomes smaller as $r$ approaches the ISCO radius. 
Comparing photon emissions from the ISCO, 
we can see that the area of the escape cone in the case of $a=0.999$ is smaller than that of $a=0.9$.
Figures~\ref{fig:cones}(g)--\ref{fig:cones}(i) show escape cones in the cases 
$r=2.98, 2, 1.001$, respectively, for $a=1$.
The area of the escape cones becomes smaller as $r$ decreases. 
It must be noted that,
according to Fig.~\ref{fig:cones}(i), 
even if the radial coordinate value of an emission point is sufficiently close to the horizon, $r=1$,
the escape cone still occupies over half of the unit 
sphere, indicating that more than half of the photons isotropically emitted 
from a circularly orbiting source can escape to infinity. In particular, 
as the entire region for which $0 \leq \alpha< \pi$ is included in the escape cone, 
all photons emitted forwardly from the source can escape to infinity.

\medskip

We now assume that photon emission is isotropic and then evaluate the escape 
probability; this 
is identified with the solid angle of an escape cone divided by $4\pi$, i.e., 
\begin{align}
P=\frac{1}{4\pi}\int_S \mathrm{d}\alpha\:\!\mathrm{d}\beta \sin \beta.
\end{align}
In subextremal cases, in terms of critical angles, 
the escape probability $P$ 
can be written as 
\begin{align}
P=1
- \frac{1}{2\pi} \int_{r_{1}^\mathrm{c}}^r \mathrm{d}r_1 
\frac{\mathrm{d}\alpha_{1(\mathrm{b})}}{\mathrm{d}r_1} \cos \beta_{1(\mathrm{b})}
- \frac{1}{2\pi} \int_{r}^{3} \mathrm{d}r_1 
\frac{\mathrm{d}\alpha_{1(\mathrm{c})}}{\mathrm{d}r_1} \cos \beta_{1(\mathrm{c})}
-\frac{1}{2\pi} \int_{3}^{r_{2}^{\mathrm{c}}}\mathrm{d}r_2 
\frac{\mathrm{d}\alpha_2}{\mathrm{d}r_2} \cos \beta_2.
\end{align}
All the integrands in the last three terms coincide with each other,
\begin{align}
\left.\frac{\mathrm{d}\alpha_{1(\mathrm{b})}}{\mathrm{d}r_1} \cos \beta_{1(\mathrm{b})}\right|_{r_1=x}
=\left.\frac{\mathrm{d}\alpha_{1(\mathrm{c})}}{\mathrm{d}r_1} \cos \beta_{1(\mathrm{c})}\right|_{r_1=x}
=\left.\frac{\mathrm{d}\alpha_2}{\mathrm{d}r_2} \cos \beta_2\right|_{r_2=x}\equiv g(x).
\end{align}
Hence, we have 
\begin{align}
P=1-\frac{1}{2\pi}\int_{r_{1}^\mathrm{c}}^{r_{2}^\mathrm{c}}g(x)\:\!\mathrm{d}x.
\end{align}
In the extremal case (i.e., $a=1$),
we can also write $P$ in terms of critical angles as follows:
\begin{align}
P=1-\frac{1}{2\pi}\int_0^{r_{\mathrm{h}}}
\mathrm{d}r_1 \frac{\mathrm{d} \alpha_{1(\mathrm{a})}}{\mathrm{d} r_1} 
\cos \beta_{1(\mathrm{a})}
-\frac{1}{2\pi} \int_{r_{\mathrm{h}}}^{r_{2}^\mathrm{c}} \mathrm{d} x \:\!g(x).
\end{align}

\begin{figure}[t!]
\centering
 \includegraphics[width=12cm,clip]{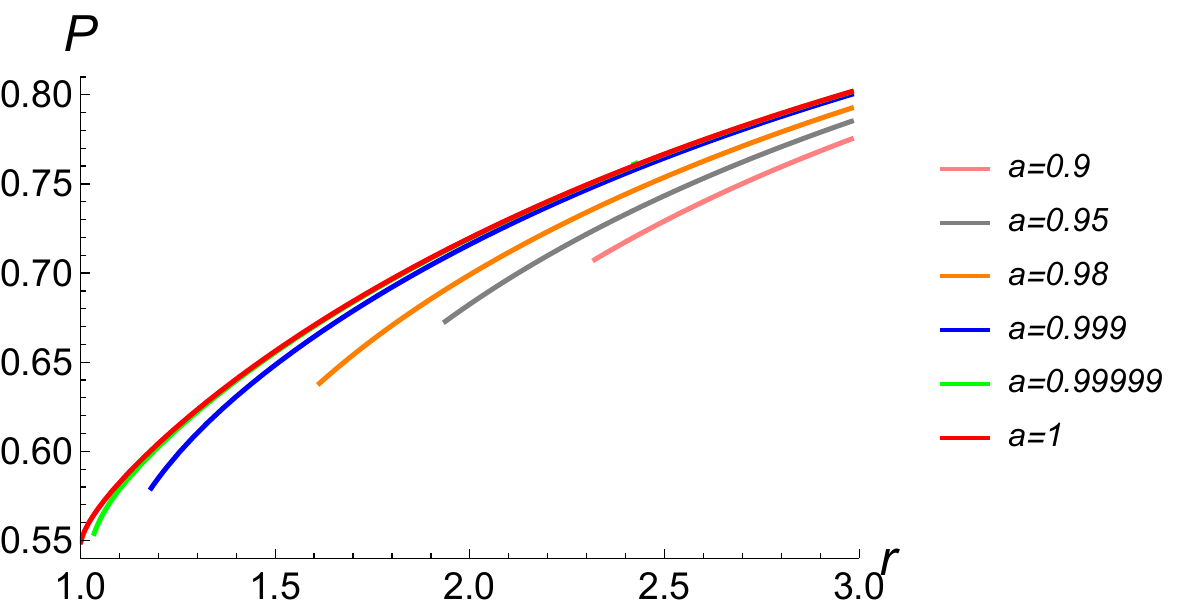}
 \caption{Dependence of the escape probability $P$ 
 on the orbital radius $r$ of a circularly orbiting source. 
 The pink, gray, orange, blue, green, and red lines correspond to the cases
 $a=0.9, 0.95, 0.98, 0.999, 0.99999$, and $1$, respectively. The left end point on each line 
 corresponds to ISCO for subextremal cases, and the left end point for $a=1$ is $r=1.001$. 
}
 \label{fig:P}
\end{figure}

Figure~\ref{fig:P} shows the dependence of the photon escape probability $P$ 
on the orbital radius $r$ of a circularly orbiting source. 
The pink, gray, orange, blue, green, and 
red lines show the cases $a=0.9, 0.95, 0.98, 0.999, 0.99999$, and $1$, respectively. 
All $P$ values decrease monotonically as $r$ decreases toward $r_{\mathrm{I}}$. 
Furthermore, the value of $P$ evaluated at $r=r_{\mathrm{I}}$ 
decreases monotonically as $a$ approaches $1$. 
For example, in the Thorne limit $a=0.998$~\cite{Thorne:1974ve}, 
the value of $P$ evaluated at the ISCO is 
\begin{align}
P(r_{\mathrm{I}})=0.5880\cdots,
\end{align}
where $r_{\mathrm{I}}=1.236\cdots$. 
These results are consistent with a naive expectation that 
$P$ becomes 
increasingly smaller as $r$ 
approaches $r_{\mathrm{h}}$. 
However, it is worth 
noting that
$P(r_{\mathrm{I}})$ does not approach zero as $a$ approaches $1$. 
Even if a photon is emitted from the source circularly orbiting 
a near-extremal Kerr black hole 
with the orbital radius $r=r_{\mathrm{I}}\simeq r_{\mathrm{h}}$, 
the escape probability is about $55 \%$. 
For the extremal case $a=1$, in the limit 
as $r$ approaches $r_{\mathrm{I}}$ (i.e., the horizon radius $r_{\mathrm{I}}=r_{\mathrm{h}}=1$),
$P$ takes a nonzero value,
\begin{align}
\lim_{r
\to 1^+}P=0.5464\cdots.
\end{align}
This indicates that more than half of photons emitted in the vicinity of the horizon escape to infinity without falling into the black hole.

\medskip

\begin{figure}[t!]
\centering
 \includegraphics[width=16.2cm,clip]{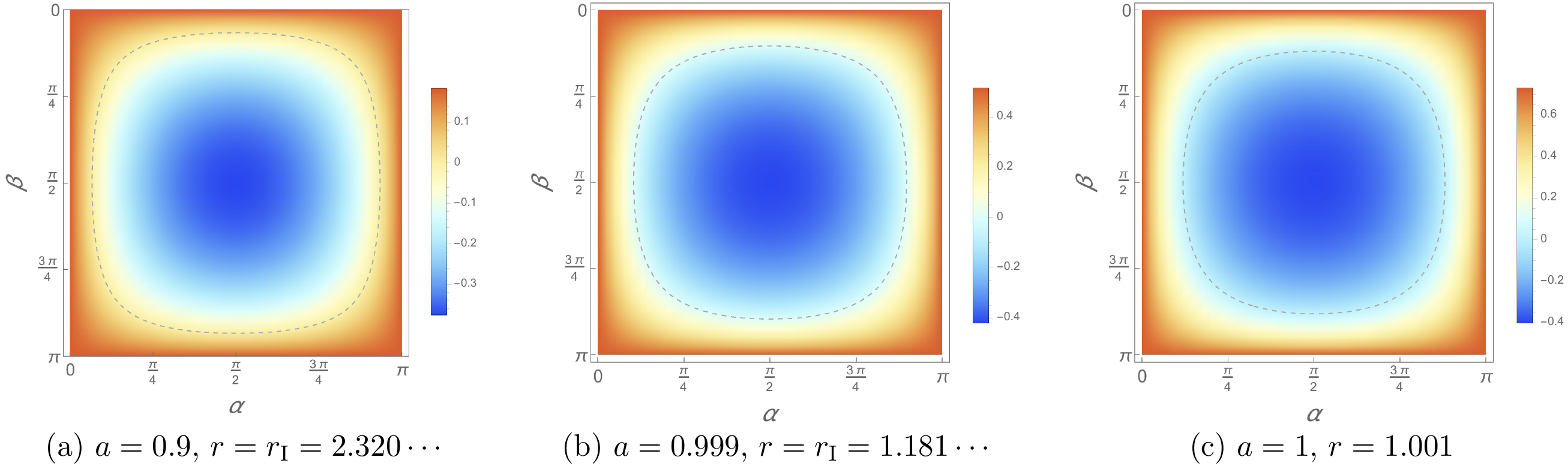}
 \caption{Density contour plot of the redshift factor $z$ in the ranges $0\leq \alpha\leq \pi$ and $0\leq \beta\leq \pi$ (i.e., 
 front side emission). 
The gray dashed lines indicate the contour $z=0$. 
The red and blue 
regions denote 
redshift and blueshift, respectively.}
 \label{fig:redshift}
\end{figure}

We now evaluate the frequency shift of 
photons escaping from the ISCO to infinity. 
The redshift factor $z$ measured by a static observer at infinity is given by 
\begin{align}
1+z=-k^{(0)}.
\end{align}
Figures~\ref{fig:redshift}(a)--\ref{fig:redshift}(c) show the density contour of $z$ 
for photons emitted from the front hemisphere of the source, 
which corresponds to the cases in 
Figs.~\ref{fig:cones}(c), \ref{fig:cones}(f), and \ref{fig:cones}(i), respectively.
The gray dashed lines show the contours of $z=0$, 
and the red/blue regions show the emission angles 
with which a photon is redshifted/blueshifted. 
The redshift factor is determined by the competition 
between the gravitational redshift and the Doppler blueshift. 
As $a$ increases, the emission angles indicating blueshift decrease, 
but more than half of the hemisphere still shows blueshift.%
\footnote{In the extremal Kerr geometry, it is known that $z$ varies smoothly with the directional 
cosine $\Psi$ with respect to $\phi$-direction, 
and there always exists the range of $\Psi$ where $z$ shows 
a net blueshift~\cite{Cunningham:1973}.}
We interpret that the Doppler blueshift due to the proper motion of the source 
overcomes the gravitational redshift.
Therefore, combined with the escape probability results, 
this implies that photons escaping from the 
vicinity of the black hole horizon certainly transport 
the information of the near-horizon region and relativistic phenomena to distant observers.

\section{Summary and discussions}
\label{sec:4}
We have considered the observability of an isotropically emitting point source 
on a circular orbit near the horizon of a rapidly rotating black hole. 
When the source in a circular orbit loses energy and angular momentum, 
it eventually reaches the ISCO. 
Because the ISCO radius is very close to the horizon radius in a near-extremal Kerr spacetime, 
the observability of the ISCO 
indicates that of the near-horizon region 
and is essential for distinguishing whether or not the central object is a black hole.

We have provided a method to calculate the escape cone of a photon emitted from the 
source and shown it explicitly; our results indicate that a photon emitted 
forward and outward from the source is likely to escape to 
infinity, while one emitted backward and inward is likely to fall into the horizon. 
On the basis of these results, we have evaluated the escape probability of a photon 
as a function of the orbital radius of the circularly orbiting source. 
As a result, we have found that it decreases monotonically 
as the radius approaches the ISCO. 
Furthermore, we have shown that the escape probability evaluated at the ISCO 
decreases monotonically when the Kerr parameter approaches the extremal value. 
Despite the fact that the ISCO radius eventually coincides with the horizon radius 
in the extremal case~\cite{Bardeen:1972fi,Kapec:2019hro}, the probability does not
become zero even in the limit of the ISCO; on the contrary, it remains at $54.6 \%$. 
In a near-extremal Kerr spacetime, the ISCO radius is always larger than 
the innermost spherical photon orbit radius (i.e., unstable circular orbit radius).
This fact implies that a photon emitted inward from the ISCO can be bounced back 
at a turning point by a potential barrier if it has a specific impact parameter. 
In the extremal Kerr spacetime, a photon can have a turning point arbitrarily close to the horizon.
Hence, photons can be scattered despite being emitted inward from a region very close to the horizon. This is why the escape probability becomes nonzero (or has nonzero measure) 
in the ISCO limit of the (near) extremal Kerr. 
We leave a more clear explanation of relatively large escape probability by the extremal Kerr throat geometry as future work.
Because the value of $54.6 \%$ is larger than $29.1 \%$---the escape probability of a photon 
emitted from the locally nonrotating source in the limit 
as the source approaches the horizon~\cite{Ogasawara:2019mir}---we can interpret this enhancement as the result of the boost (or the relativistic beaming) 
due to the circular motion of the source. 
We have also found that photons emitted from the front side of the source 
get blueshifted for distant observers, i.e., 
the Doppler blueshift due to the proper motion of the source can overcome the gravitational redshift.

On the basis of these results, 
we conclude that the ISCO of a rapidly rotating black hole spacetime is observable. 
Our results provide insight into optical phenomena 
such as the observation of a black hole shadow. 
The ISCO is often identified as the innermost edge of the accretion disk or 
is considered as the position where the phase of accretion flow switches; 
however in any case, the orbit is located 
in the immediate vicinity of the near-extremal rotating black hole. 
Photons emitted from this region contain information 
on the near-horizon region and phenomena, and certainly transport it to infinity.
Finally, the information can be obtained from photons 
that appear as the edge of the black hole shadow to a distant observer.
From this, we can conclude that if a black hole is rapidly spinning,
signs of near-horizon physics will be detectable on the edge of the shadow.

\begin{acknowledgments}
The authors thank T.~Harada, H.~Ishihara, U.~Miyamoto, and M.~Takahashi for useful comments. 
They also especially thank the anonymous referee for reading our manuscript carefully, checking our calculations, and providing many helpful comments and suggestions.
This work was supported by Grant-in-Aid for Early-Career Scientists~(JSPS KAKENHI Grant No.~JP19K14715) (T.I.), Grant-in-Aid for JSPS Fellows~(JSPS KAKENHI Grant No.~JP18J10275) (K.O.) from the Japan Society for the Promotion of Science, and 
Rikkyo University Special Fund for Research (K.N.). 

\end{acknowledgments}

\end{document}